\documentclass[
 reprint,
 amsmath,amssymb,
 aps,
 nolongbibliography
]{revtex4-2}

\usepackage{xcolor}
\usepackage{graphicx}
\usepackage{dcolumn}
\usepackage{bm}

\begin{document}

\preprint{APS/123-QED}

\title{\textcolor{black}{Frequency correlation requirements on the biphoton wavefunction in an induced coherence experiment between separate sources}}

\author{Arturo Rojas-Santana}
\email{jorge-arturo.rojas@icfo.eu}
\affiliation{ICFO - Institut de Ciencies Fotoniques, The Barcelona Institute of Science and Technology, 08860 Castelldefels (Barcelona), Spain}
\affiliation{Tecnologico de Monterrey, Escuela de Ingenier\'ia y Ciencias, Ave. Eugenio Garza Sada 2501, Monterrey, N.L. 64849, Mexico}

\author{Gerard J. Machado}
\affiliation{ICFO - Institut de Ciencies Fotoniques, The Barcelona Institute of Science and Technology, 08860 Castelldefels (Barcelona), Spain}

\author{Dorilian Lopez-Mago}
\affiliation{Tecnologico de Monterrey, Escuela de Ingenier\'ia y Ciencias, Ave. Eugenio Garza Sada 2501, Monterrey, N.L. 64849, Mexico}

\author{Juan P. Torres}
\affiliation{ICFO - Institut de Ciencies Fotoniques, The Barcelona Institute of Science and Technology, 08860 Castelldefels (Barcelona), Spain}
\affiliation{Departament of Signal Theory and Communications, Universitat Politecnica de Catalunya, 08034 Barcelona, Spain}

\date{\today}

\begin{abstract}
There is renewed interest in using the coherence between beams generated in separate down-converter sources for new applications in imaging, spectroscopy, microscopy and optical coherence tomography (OCT). These schemes make use of continuous wave (CW) pumping in the low parametric gain regime, \textcolor{black}{which produces frequency correlations, and frequency entanglement, between signal-idler pairs generated in each single source. But can induced coherence still be observed if there is no frequency correlation, so the biphoton wavefunction is factorable?} We will show that this is the case, and this might be an advantage for OCT applications. High axial resolution requires a large bandwidth. For CW pumping this requires the use of short nonlinear crystals. This is detrimental since short crystals generate small photon fluxes. We show that the use of ultrashort pump pulses allows improving axial resolution even for long crystal that produce higher photon fluxes.
\end{abstract}


\maketitle


\section{\label{sec:level1}Introduction}
In 1991 Zou et al. \cite{Zou1991,zou1992} demonstrated that indistinguishability of idler beams generated at separate parametric down-converting sources can {\em induce coherence} between the signal beams generated at the same separate sources. The effect was demonstrated originally at the low parametric gain regime, where the probability to generate pairs of photons simultaneously in each source is very low. However coherence is also observed in the high parametric gain regime \cite{belinsky1992,wiseman2000}. Here we are interested in the low parametric gain regime, since this scenario allows to quantify the degree of entanglement between down-converted photons straightforwardly.   

Induced coherence in a system of two parametric down-converters is a particular case of a broader class of interferometers sometimes referred as {\em nonlinear interferometers} \cite{Chekhova2016} based on optical parametric amplifiers. The last few years have seen a surge of interest in using these interferometers for new schemes in imaging \cite{Barreto(2014),cardoso2018}, sensing \cite{kutas2020}, spectroscopy \cite{kalasnikov2016,Paterova2018}, microscopy \cite{Kviatkovsky2020,paterova2020} and OCT \cite{shapiro2009,valles2018optical,Paterova2018OCT,vanselow2019}.  One advantage of these systems is that one can choose a wavelength for the beam that interacts with the sample and is never detected, and another wavelength for the beam to be detected that enhance photo-detection efficiency. They also can show better sensitivity than alternative schemes \cite{dowling2019,miller2020}.

Up to now all experiments but two \cite{shapiro2009,gerard2020} are performed in the low parametric gain regime. In all these cases the bandwidth of the pump laser ($\delta_p$) is considerably smaller than the bandwidth of down-conversion ($\Delta_{dc}$) \cite{dayan2007}. This produces a high degree of entanglement between signal and idler beams generated in a single biphoton source. This can lead to think that frequency entanglement between signal-idler pairs generated in a nonlinear crystal is a necessary condition to observe induced coherence between signal photons generated in separate nonlinear crystals. We will demonstrate below that induced coherence happens when there is no frequency correlation, and thus no frequency entanglement, so importantly CW pumping is not a requisite to observe induced coherence. As we will show, this can have important practical consequences for the implementation of high-flux and high-resolution optical coherence schemes based on induced coherence.

\textcolor{black}{When we consider only frequency correlations between signal-idler pairs, the quantum state can be described by the biphoton wavefunction $\Phi(\omega_s,\omega_i)$, where $\omega_{s,i}$ refer to the frequency of signal and idler photons, respectively. If the biphoton function is factorable, i. e. $\Psi(\omega_s,\omega_i)=F(\omega_s) G(\omega_i)$ the state is separable and shows no frequency entanglement. If the state cannot be decomposed in this way, the quantum state is entangled. For pure states, the entropy of entanglement \cite{parker2000} is a good quantitative measure of how much entanglement there is between signal and idler photons generated. It is large when the ratio $\Delta_{dc}/\delta_p \gg 1$ or $\Delta_{dc}/\delta_p \ll 1$. The degree of entanglement can also be retrieved from the number of modes present in the Schmidt decomposition of the quantum state \cite{law2000}.}

\textcolor{black}{Frequency entanglement in parametric down-conversion has been analyzed under different circumstances \cite{keller1997,grice2001}. Several methods to tailor the frequency correlations, and thus the degree of entanglement between signal-photon pairs, has been proposed and demonstrated. Signal-idler pairs that show frequency correlation, in contrast to the frequency anti-correlation that arises under CW pumping, has been produced \cite{kuzucu2005}, as well as paired photons in a separable state \cite{alejandra2007,martin2008,mosley2008}. Certain methods even allow to generate any type of frequency correlation between signal-idler pairs \cite{aop2010}, as well as tailoring the bandwidth of the down-converted photons \cite{martin2009}.}

\begin{figure}[t!]
\centering\includegraphics[width=8.5 cm]{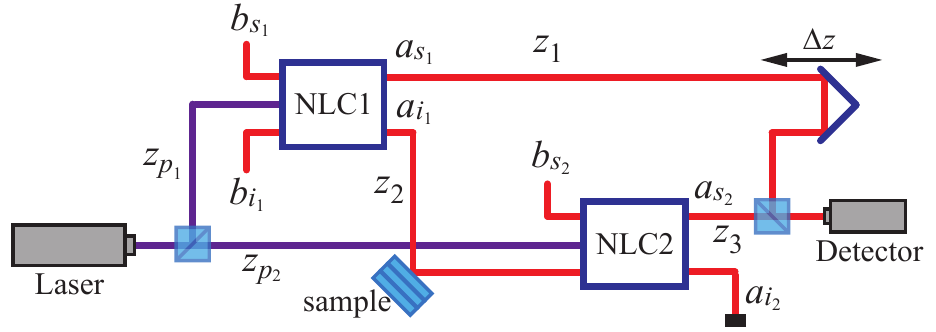}
\caption{Induced coherence between signal photons generated in separate parametric down-converters. The idler traverses a lossy sample before being injected into NLC$_2$. The detector measures the interference between signal photons $s_{1}$ and $s_{2}$ as a function of the path delay $\Delta z$. NLC: nonlinear crystal; $s$,$i$: signal and idler modes; $b$, $a$ input/output quantum operators.} \label{Fig:1}
\end{figure}

\section{Role of signal-idler entanglement for observing induced coherence}
Figure~\ref{Fig:1} shows a scheme of an induced coherence experiment with two parametric down-converters (NLC$_1$ and NLC$_2$). We consider a pulsed laser that generates coherent light with a spectrum $F(\Omega_{p})$. The frequency of the pump is $\omega_p=\omega_p^0+\Omega_p$, with $\omega_{p}^0$ being the central frequency and $\Omega_{p}$ the frequency deviation from the central frequency. A beam splitter divides the pump beam into two coherent sub-beams that pump the two nonlinear crystals. The two sub-beams travel distances $z_{p_1}$ and $z_{p_2}$ before reaching NLC$_1$ and NLC$_2$, respectively.

Both crystals have nonlinear susceptibility $\chi^{(2)}$ and length $L$. The nonlinear interaction generates signal and idler photons $s_1$ and $i_1$ in NLC$_1$, and $s_2$ and $i_2$ in NLC$_2$. The frequency of the signal and idler photons reads $\omega_{s}=\omega_s^0+\Omega_s$ and $\omega_{i}=\omega_i^0+\Omega_i$, where $\omega_{s,i}^0$ are central frequencies and $\Omega_{s,i}$ are frequency deviations from the corresponding central frequencies. The conditions $\omega_{p}^0=\omega_{s}^0+\omega_{i}^0$ and $\Omega_{p}=\Omega_{s}+\Omega_{i}$ are satisfied.  

The quantum operators $a_{s_1,s_2}(\Omega_s)$ and $a_{i_1,i_2}(\Omega_i)$ correspond to signal and idler modes at the output face of the corresponding nonlinear crystals. $b_{s_1,s_2}(\Omega_s)$ and $b_{i_1}(\Omega_i)$ designate the corresponding operators at the input face. In the low parametric gain regime, the Bogoliubov transformations that relate the input-output operators for NLC$_1$ are \cite{banaszek2006,Torres2011}:
\begin{eqnarray}
&&a_{s_1} (\Omega_{s}) = U_s(\Omega_s) b_{s_1} (\Omega_{s})+\int \mathrm{d} \Omega_{i} V_{s_1} (\Omega_{s},\Omega_i) b_{i_1}^{\dagger} (\Omega_{i}),~~~~~\label{Equation1}\\
&&a_{i_1} (\Omega_{i}) = U_i(\Omega_i) b_{i_1} (\Omega_{i}) + \int \mathrm{d}\Omega_s V_{i_1}(\Omega_{s},\Omega_i) b_{s_1}^{\dagger} (\Omega_{s}),~~~\label{Equation1b}
\end{eqnarray}
where $U_s(\Omega_s)=\exp\left[ ik_{s}(\Omega_{s}) L \right]$, $U_i(\Omega_i)=\exp\left[ ik_{i}(\Omega_i) L \right]$ and
\begin{eqnarray}
& & V_{s_1} (\Omega_{s},\Omega_i) = i (\sigma L) F_{p_1}(\Omega_s+\Omega_i)  \mathrm{sinc}\left[ \frac{ \Delta k L}{2} \right] \nonumber \\
& & \times \exp\left[i\frac{k_{p}(\Omega_{s}+\Omega_i)+k_{s}(\Omega_{s}) - k_{i}(\Omega_{i})}{2}L  \right],    \label{Vs1}\\
& & V_{i_1} (\Omega_{s},\Omega_i) =i(\sigma L)  F_{p_1}(\Omega_s+\Omega_i) \mathrm{sinc}\left[ \frac{ \Delta k L}{2} \right] \nonumber \\
& & \times \exp\left[i\frac{k_{p}(\Omega_{s}+\Omega_i)+k_{i}(\Omega_{i}) - k_{s}(\Omega_{s})}{2}L  \right]. 
\end{eqnarray}
The nonlinear coefficient $\sigma$ is \cite{banaszek2006,dayan2007,Torres2011}
\begin{equation}
    \sigma = \left[ \frac{\hbar \omega_{p}^{0} \omega_{s}^{0}\omega_{i}^{0} [\chi^{(2)} ]^{2} N_{0} }{16\pi \epsilon_{0} c^{3} n_{p} n_{s} n_{i} A}  \right]^{1/2},
\end{equation}
where $N_0$ is the number of pump photons per pulse, $A$ is the effective area of interaction and $n_{p,s,i}$ are refractive indexes at the central frequencies of all waves involved. The function $F_{p_1}$ is
\begin{equation}
   F_{p_1}(\Omega_p) = \frac{T_{0}^{1/2}}{\pi^{1/4}}\exp\left[ - \frac{\Omega_p^2 T_0^2}{2} \right]  \mathrm{exp}\left[ i k_p(\Omega_{p}) z_{p_1} \right],
\label{Eq:Fp12}
\end{equation}
where we have assumed a Gaussian shape for the spectrum of the pump beam. The function $F_p$ is normalized to $1$. $T_0$ is the temporal width of the pump pulses. The wave-vector phase mismatch is $\Delta k=k_p(\Omega_s+\Omega_i)-k_s(\Omega_s)-k_i(\Omega_i)$. If we expand in Taylor series to first order the wave-vectors as $k_i(\Omega)=k_i^0+N_i \Omega$ ($N_{p,s,i}$ are inverse group velocities) and assume perfect phase matching at the central frequencies ($k_p^0=k_s^0+k_i^0$), we obtain $\Delta k=D_{+} \Omega_p+D \Omega_{-}/2$, where $\Omega_{-}=\Omega_s-\Omega_i$, $D_{+}=N_p-(N_s+N_i)/2$ and $D=N_i-N_s$.

The idler mode $a_{i1}$ traverses a distance $z_2$ before encountering a lossy sample characterized by reflectivity $r(\Omega_i)$. The quantum operator transformation that describes this process is \cite{haus2000,Boyd2008}
\begin{equation}
    a_{i_1}(\Omega_i) \longrightarrow  r(\Omega_i) a_{i_1}(\Omega_i) \exp \left[ i k_i(\Omega_i)z_2 \right] +f(\Omega_i), \label{Eq:sample}
\end{equation}
where the operator $f$ fulfills the commutation relationship $[f(\Omega),f^{\dagger}(\Omega^{\prime})]=(1-|r(\Omega)|^2)\delta(\Omega-\Omega^{\prime})$. 

The idler beam is injected into NLC$_2$ so that the operator $a_{s_2}$ that describes signal beam $s_2$ at the output face of NLC$_2$ is
\begin{eqnarray}
    & & a_{s2}(\Omega_{s}) = U_s(\Omega_s) b_{s_2}(\Omega_{s})  +\int \mathrm{d} \Omega_{i} V_{s_2}(\Omega_s,\Omega_i) f^{\dagger}(\Omega_i) \label{Equation2} \\
    & & + \int \mathrm{d} \Omega_{i} r^*(\Omega_i)V_{s_2}(\Omega_s,\Omega_i) U_i^*(\Omega_i) \exp\left[-i k_i(\Omega_i) z_2 \right] b_i^{\dagger}(\Omega_i), \nonumber
\end{eqnarray}
where only terms up to first order in $\sigma L$ has been considered, the terms that give a non-zero contribution in the calculation of the first-order correlation function. The function $V_{s_2}$ is analogous to $V_{s_1}$ in Eq. (\ref{Vs1}) with $F_{p_2}=F_p(\Omega_p) \exp \left[ i k_p(\Omega_p) z_{p_2} \right]$.

Signal photon $s_1$ traverses a distance $z_1$ before detection, and signal photon $s_2$ traverses a distance $z_3$. The number of signal photons generated per pulse, $N_{s_1}=\int \mathrm{d}\Omega\, a_{s_1}^{\dagger}(\Omega)a_{s_1}(\Omega)$ and $N_{s_2}=\int \mathrm{d}\Omega\, a_{s_2}^{\dagger}(\Omega)a_{s_2}(\Omega)$ is 
\begin{equation}
    N_{s_1}=N_{s_2}=2\pi \frac{\sigma^{2} L}{D}.
    \label{Equation3}
\end{equation}
It depends on the total number of pump photons per pulse, however it is independent of the shape of the pulse. This fact and that $N_{s_1}=N_{s_2}$ are characteristics of the low parametric gain regime.

We are interested in the normalized first-order correlation function $g_{s_1,s_2}^{(1)}$ between beams $s_1$ and $s_2$ that gives the visibility of the interference fringes detected after combining both signals in a beam splitter, i.e.,
\begin{equation}
    g_{s_1,s_2}^{(1)}=\frac{1}{N_{s_1}^{1/2} N_{s_2}^{1/2} }\int \mathrm{d}\Omega\, a_{s_1}^{\dagger}(\Omega)a_{s_2}(\Omega).
    \label{Equation4}
\end{equation}
 Let us first assume that there are no losses in the idler path ($r(\Omega)=1$). Using Eqs.~(\ref{Equation1}), (\ref{Equation2}) and (\ref{Equation3}) into Eq.~({\ref{Equation4}}) and taking into account the distances $z_1$ and $z_3$ that signal beams $s_1$ and $s_2$ propagate before combination in the beam splitter, the first-order correlation function can be written as
\begin{eqnarray}
 & & \left |g_{s1,s2}^{(1)}(T_1,T_2)\right | = \mathrm{tri}\left(\frac{T_1}{DL}\right) \nonumber \\
 & & \times \exp\bigg[- \frac{1}{16T_{0}^{2}} \left[\left(1-\frac{2D_{+}}{D}\right)T_1+2 T_2\right]^{2}\bigg], \label{gs1}
\end{eqnarray}
where $\mathrm{tri}(\xi/2)=1/\pi \,\int \mathrm{sinc}^2(x) \mathrm{exp}(i\xi x)\mathrm{d}x  $ is the triangular function and
\begin{eqnarray}
& & T_1 = \frac{z_3-z_1+z_2}{c}+N_i L,  \label{Eq:T1} \\
& & T_2 = \frac{z_{p_2}-z_{p_1}-z_2}{c}-N_i L. \label{Eq:T2}
\end{eqnarray} 
We assume that the condition $z_{p2}=z_{p1}+cN_i L+z_2$ is fulfilled, so that $T_2=0$. In order to optimize pulsed parametric amplification in NLC$_2$ one needs to synchronize the time of arrival of pump and idler pulses to the nonlinear crystal \cite{shapiro2009}.

The first-order correlation function is the product of a triangular function of width $DL$ and a Gaussian function of width $T_0$.  Figure 2 plots the first-order correlation function as a function of $\Delta z=z_3-z_1+z_2+c N_i L$ for a crystal length $L=5$ mm and three different pulse widths: $T_{0}=100$~ps, $T_{0}=2$~ps and $T_{0}=100$~fs.  $\Delta z$ can be modified in an experiment by changing the pathlength difference $z_3-z_1$. We have considered as example two MgO-doped LiNbO$_3$ crystals \cite{MgO} pumped by a pulsed laser operating at $\lambda_{p}^{0}=532$ nm. The resulting type-0 signal and idler beams have wavelengths $\lambda_{s}^{0}=810$ nm and $\lambda_{i}^{0}=1550$ nm with $D=-263.50$~fs/mm and $D_{+}=780$~fs/mm.

In the limiting case of CW pumping ($T_0 \rightarrow \infty$), the shape of the first-order correlation function is dominated by the triangular function [see Fig. 2(a)],  as it has been measured in many occasions \cite{valles2018optical}. As we decrease the temporal width of the pump pulses, the influence of the triangular and Gaussian functions on $g_{s_1,s_2}^{(1)}$ becomes comparable [Fig. 2(b)].  Finally, when $T_0 \ll DL$, the shape of the first-order correlation function is dominated by the Gaussian function [Fig. 2(c)].  

\begin{figure}[t!]
\centering\includegraphics[width=8.5cm]{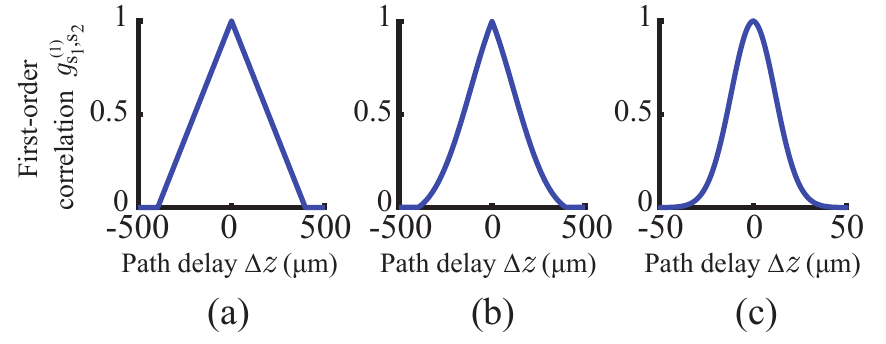}
\caption{First-order correlation function as a function of the path delay $\Delta z$. We consider a crystal with length $L=5$ mm. The pump pulses have temporal widths: (a) $T_{0}=100$~ps; (b) $T_{0}=2$~ps; and (c) $T_{0}=100$~fs.} \label{Fig:2}
\end{figure}

Is entanglement between signal and idler photons relevant for observing induced coherence? Inspection of Fig. 2 shows that it is not, since for all values of $T_0$ and crystal length $L$, that correspond to quantum states with different degrees of entanglement, there is induced coherence. For the sake of clarity, let us be more specific. In the low parametric gain regime, the biphoton function
\begin{equation}
 \Psi(\Omega_s,\Omega_i) =i\sigma L F(\Omega_s+\Omega_i) \mathrm{sinc}\left[ \frac{ \Delta k L}{2} \right] \exp\left(i s_k L  \right),
 \label{biphoton}
\end{equation} 
where $s_k=k_p(\Omega_s+\Omega_i)+k_s(\Omega_s)+k_i(\Omega_i)$, determines the nature of the correlations between the paired photons and the degree of entanglement between them \cite{Torres2011}. If we can decompose $\Psi(\Omega_s,\Omega_i)$ into two functions that depend separately on the variables $\Omega_s$ and $\Omega_i$ the quantum state is non-entangled (separable).

For the sake of simplicity, let us consider $D_+=0$ and make the approximation  $\text{sinc}(x) \sim \exp(-\alpha^2 x^2)$ with $\alpha=0.455$ \cite{clara2008}. The normalized biphoton function derived from Eq. (\ref{biphoton}) is
\begin{eqnarray}
 & & \Phi(\Omega_s,\Omega_{i})= \left( \frac{\alpha T_0 DL}{\sqrt{2} \pi} \right)^{1/2} \exp\left[ - \frac{(\Omega_s+\Omega_i)^2 T_0^2}{2} \right] \nonumber \\
 & & \times \exp\left[- \frac{\alpha^2 (D L)^2}{16} (\Omega_s-\Omega_i)^2 \right]. \label{Phi}
\end{eqnarray}
$|\Phi(\Omega_s,\Omega_i)|^2$ yields the probability to detect a signal photon at frequency $\omega_s^0+\Omega_s$ in coincidence with an idler photon at frequency $\omega_i^0+\Omega_i$. 

\begin{figure}[t!]
\centering\includegraphics[width=8.5 cm]{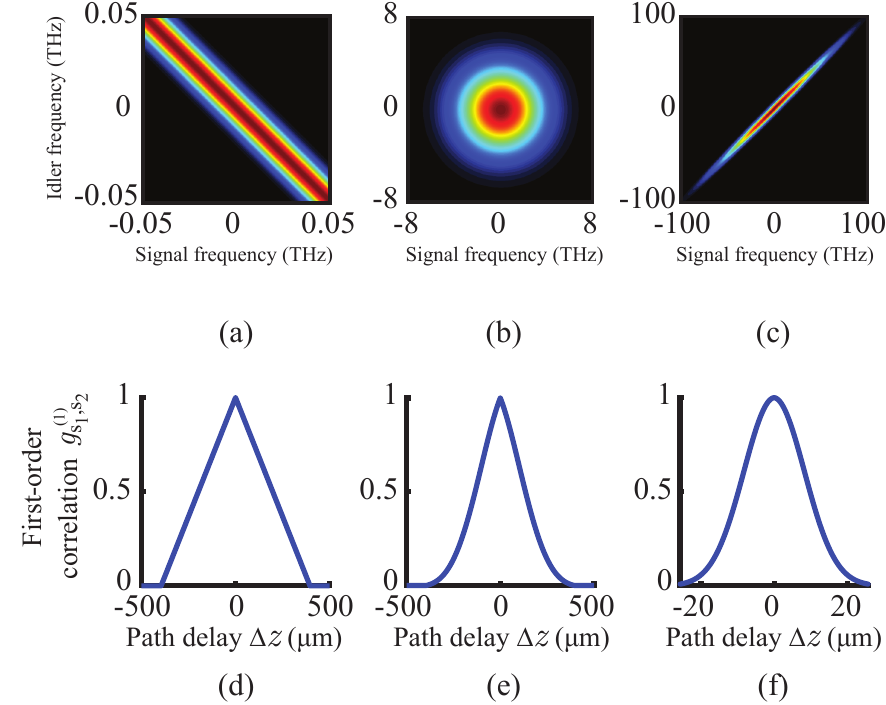}
\caption{(a), (b) and (c): Normalized  biphoton  function $|\Phi(\Omega_s,\Omega_i)|^2$. The axis correspond to angular frequency deviation $\Omega_s$ and $\Omega_i$. (d), (e) and (f): First-order correlation function.  The pump pulse durations $T_0$ are: (a) and (d) $T_{0}=100$~ps; (b) and (e) $T_{0}=212$~fs; (c) and (f) $T_{0}=10$~fs. The nonlinear crystal length is $L=5$~mm.} \label{Fig:3}
\end{figure}

The degree of entanglement depends on the ratio between the bandwidth of the pump beam and the bandwidth of down-conversion: $\gamma=\alpha D L/(2\sqrt{2} T_0)$. \textcolor{black}{For $\gamma=1$ we can write the quantum state as $\Phi(\Omega_s,\Omega_i)=\Phi_s(\Omega_s) \Phi_i(\Omega_i)$, the state is separable}. The degree of entanglement is high if $\gamma \gg 1$ or $\gamma \ll 1$ \cite{parker2000,hendrych2009}. Figures \ref{Fig:3}(a), (b) and (c) plot $|\Phi(\Omega_s,\Omega_i)|^2$ for a crystal length $L=5$ mm and three different pump pulse widths that correspond to $\gamma \ll 1$ ($T_{0}=100$~ps), $\gamma=1$ ($T_{0}=212$~fs) and $\gamma \gg 1$ ($T_{0}=10$~fs). 

When $T_0 \gg DL$ [Fig. \ref{Fig:3}(a)] there is frequency anti-correlation between signal and idler photons. One can detect coincidences if $\Omega_i \sim -\Omega_s$. For $T_0 \ll DL$ [Fig. \ref{Fig:3}(c)] there is frequency correlation, there are coincidences only if $\Omega_i \sim \Omega_s$. In between, the degree of correlation is low and the quantum state can become separable [Fig. \ref{Fig:3}(b)].  Figures \ref{Fig:3}(d), (e) and (f) show the first-order correlation function corresponding to these cases. For all values of the degree of entanglement we observe coherence, ruling out that the entanglement nature of the paired photons is responsible for the existence of induced coherence.

\section{Optical coherence tomography with large bandwidth and high photon flux}
OCT is an optical imaging technique that permits cross-sectional and axial high-resolution tomographic imaging \cite{Huang1991}. The axial and transverse resolutions are independent. To obtain information in the axial direction (along the beam propagation), OCT uses a source of light with large bandwidth that allows optical sectioning of the sample. 

Different OCT schemes that make use of biphoton sources have been demonstrated. In all cases one photon of the pair probe the sample. Some schemes measure the second-order correlation function of signal and idler photons \cite{Nasr2003,Lopez-Mago2012e}, others measure the first-order correlation function of signal photons generated in different biphoton sources \cite{shapiro2009,valles2018optical} and others measure the flux of signal photons generated in an SU(1,1) nonlinear interferometer \cite{Paterova2018OCT,cardoso2018}.

Figures \ref{Fig:2} and \ref{Fig:3} demonstrate that one can observe induced coherence independently of the degree of entanglement between signal and idler beams. This has an important consequence for the further development of OCT based on nonlinear interferometers. Equation (\ref{Equation3}) shows that the photon flux generated increases with the nonlinear crystal length. However, for CW pumping, $\Delta_{dc}$ goes as $\sim 1/DL$. OCT with high axial resolution requires a large bandwidth. Therefore high axial resolution implies the generation of low photon fluxes and so longer integration times to obtain high-quality images. This is detrimental for OCT applications.

The first-order correlation function is the measure of axial resolution in an OCT system. Equation (\ref{gs1}) shows that one can obtain a narrow first-order correlation function, and thus high axial resolution, even for long nonlinear crystal by using an ultrashort pump pulse. 

In order to show this effect, we consider a bilayer sample characterized by a reflectivity $r(\Omega) = r_{0} + r_{1} \exp[i (\omega^0+\Omega) \tau]$.  The delay is $\tau=2 d_0 n_0/c$ where $d_0$ and $n_0$ designate the thickness and refractive index, respectively, of the sample. The coefficient $r_{0}$ is the Fresnel coefficient for the first layer, whereas $r_{1}$ is the effective coefficient for the second layer, taking into account propagation through the sample. $z_2$ is the distance traveled by the idler beam reflected from the first layer, while $z_2+2 n_0 d_0$ is the optical distance traveled by the idler beam reflected from the second layer. 

The signal detected at one output port of the beam splitter is
\begin{widetext}
\begin{eqnarray}
    N&=N_{s_1} \left\{ 1+ r_0 g_{s_1,s_2}^{(1)}(T_1,T_2) \sin \left[ (\omega_p^0/c) (z_{p_2}-z_{p_1})  \right. \right. \nonumber \left. -(\omega_i^0/c) (z_2+n_iL)-(\omega_s^0/c)(z_1-z_3) \right] \nonumber\\ 
    & + r_1  g_{s_1,s_2}^{(1)}(T_1^{\prime},T_2^{\prime}) \sin \left[ (\omega_p^0/c) (z_{p_2}-z_{p_1}) \right. \left. \left. -(\omega_i^0/c) (z_2+n_i L+2n_0 d_0)-(\omega_s^0/c)(z_1-z_3) \right] \right\},  \label{bilayer}
\end{eqnarray}
\end{widetext}
where $T_1^{\prime}=T_1+\tau$ and $T_2^{\prime}=T_2-\tau$. $T_1$ and $T_2$ are given by Eqs.~(\ref{Eq:T1}) and (\ref{Eq:T2}). We can choose $z_{p_2}=z_{p_1}+c N_i L+z_2$.

\begin{figure}[t!]
\centering\includegraphics[width=8.5 cm]{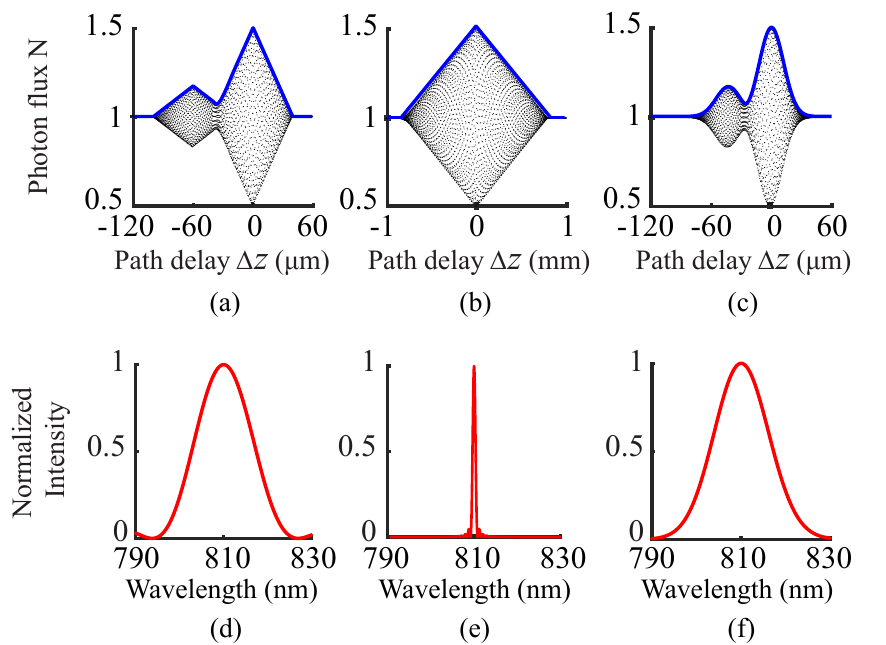}
\caption{(a), (b) and (c): Signal $N$ in one output port of the beam splitter as a function of $\Delta z$. (a) $L=0.5$~mm, $T_{0}=100$~ps, (b) $L=10$~mm, $T_{0}=100$~ps, and (c) $L=10$~mm, $T_{0}=100$~fs. (d), (e) and (f): Normalized spectrum of the signal photon. The bandwidths (FWHM) are $14.8$~nm, $0.8$~nm and $20$~nm.} \label{Fig:4}
\end{figure}

Figure~\ref{Fig:4} shows the photon flux $N$ as a function of $\Delta z$ [Eq.~(\ref{bilayer})] for a $20$~$\mu$m glass slab (refractive index $n_0=1.5$) embedded between air ($n_{1}=1$) and water ($n_{2}=1.3$). We consider three scenarios. Fig.~\ref{Fig:4}(a) considers a pump beam with $T_{0}=100$~ps (quasi CW) and a crystal with $L=0.5$~mm. The interferogram shows two maxima separated by $60$~$\mu$m, the sample's optical path length $c\tau$. 

Figure~\ref{Fig:4}(b) considers the same pulse duration but $L=10$~mm. The interferogram cannot resolve the thickness of the sample, there is not enough axial resolution. Figure~\ref{Fig:4}(c) considers the same length $L=10$~mm but now with $T_{0}=100$~fs. The interferogram recovers the two maxima, thereby resolving the layers of the sample.  The two maxima are separated by $42$~$\mu$m, which is smaller than the sample's optical thickness. This result can be understood noticing that the peak of the interferogram when the shape of the first-order correlation function is dominated by the Gaussian function will take place for a value of $T_1$ [see Eq. (\ref{gs1})]
\begin{eqnarray}
 & & \left( 1-\frac{2D_+}{D} \right) \left( T_1+\tau \right)-2 \tau=0,  \nonumber \\
 & & \Longrightarrow T_1=\frac{D + 2 D_+}{D-2D_+} \tau.
\end{eqnarray} 
Taking into account the values of $D=-263$ fs/mm and $D_+=780$ fs/mm, the factor $(D+2 D_+)/(D-2D_+)=-0.71$. The separation between the two maxima corresponding to the two layers is  $-0.71 \times 60 \mu$m $\sim -42 \mu$m. This result is reminiscent of the fact that after reflection from the sample, we have two pulses separated by $\tau$ that are injected in the second nonlinear crystal and both show certain delay with the pump pulse \cite{Yoon-HoKim2000}. For a case with $D_+=0$ we would have again $T_1=\tau$ as in the quasi CW case.

Figure~\ref{Fig:4} also shows the signal spectrum for each case, given by $S(\Omega_s)=\int \mathrm{d}\Omega_i\, |\Phi(\Omega_s,\Omega_i)|^2$. The interferograms and spectra show the reciprocal relation between spectral bandwidth and axial resolution. 

\section{Conclusions}
We have demonstrated that induced coherence between signal beams generated in separate biphoton sources can be observed independently of the degree of entanglement between signal-idler photon pairs \textcolor{black}{generated in the same nonlinear crystal}. In the first demonstration of OCT based on parametric down-conversion, in the high parametric gain regime, the bandwidth of the pump pulse and the bandwidth of down-conversion ($0.36$ nm) are made comparable due to the use of narrowband filters. However in the high parametric gain one cannot readily quantify signal-idler entanglement. 

In the low parametric gain regime, the emission rate of photon pairs increases with the length of the nonlinear crystal, regardless of the duration of the pump pulse. We have shown that an OCT scheme based on induced coherence can achieve high axial resolution and high photon emission rates by combining ultrashort pumping with long crystals. The method maintains its salutary features, i.e., probing the sample with photons centered at the most appropriate wavelength while using the optimum wavelength for photodetectors.

\begin{acknowledgments}
We acknowledge financial support from the Spanish Ministry of Economy and Competitiveness through the “Severo Ochoa” program for Centres of Excellence in R$\&$D (SEV-2015-0522), from
Fundació Privada Cellex, from Fundació Mir-Puig, and from Generalitat de Catalunya through the CERCA program. GJM was supported by the Secretaria d'Universitats i Recerca del Departament d'Empresa i Coneixement de la Generalitat de Catalunya, as well as the European Social Fund (L'FSE inverteix en el teu futur)—FEDER. DLM acknowledges funding from Consejo Nacional de Ciencia y Tecnolog\'{i}a (293471, 295239, APN2016-3140) and ARS acknowledges support from Becas de Movilidad CONACYT, SEGIB and Fundaci\'{o}n Carolina.
\end{acknowledgments}

\nocite{*}

\bibliography{apssamp}

\end{document}